\documentclass[prd,
preprint,
superscriptaddress,showpacs,nofootinbib,%
tightenlines ]{revtex4}
\usepackage{epsfig}
\usepackage{amssymb}

\newcommand{\ben}{\begin{displaymath}}
\newcommand{\een}{\end{displaymath}}
\newcommand{\be}{\begin{equation}}
\newcommand{\ee}{\end{equation}}
\newcommand{\bea}{\begin{eqnarray}}
\newcommand{\eea}{\end{eqnarray}}
\begin{document}
\title{
Properties of effective massive Yang-Mills theory in the limit of vanishing vector boson mass}
\author{J.~Gegelia}
 \affiliation{Institut f\"ur Theoretische Physik II, Ruhr-Universit\"at Bochum,  D-44780 Bochum,
 Germany}  \affiliation{Tbilisi State  University,  0186 Tbilisi,
 Georgia}
\author{U.-G.~Mei\ss ner}
 \affiliation{Helmholtz Institut f\"ur Strahlen- und Kernphysik and Bethe
   Center for Theoretical Physics, Universit\"at Bonn, D-53115 Bonn, Germany}
 \affiliation{Institute for Advanced Simulation, Institut f\"ur Kernphysik
   and J\"ulich Center for Hadron Physics, Forschungszentrum J\"ulich, D-52425 J\"ulich,
Germany}

\date{26 September, 2014}

\begin{abstract}

Two-loop corrections to the pole mass of the vector boson and the pole masses and the magnetic moments of fermions are
calculated in the framework of an effective field theory of massive Yang-Mills fields interacting with fermions.
It is shown that the limit of vanishing vector boson mass is finite for all these quantities.
Implications of the obtained results are discussed.

\end{abstract}



\pacs{03.70.+k,11.25.Db,11.15.-q}


\maketitle

\section{Introduction}

    While the Standard Model (SM) is obtained by requiring renormalizability, the modern point of
view treats it as an effective field theory
(EFT) \cite{Weinberg:mt}. In EFT divergences of loop diagrams
are removed by redefining an finite number of parameters of the effective Lagrangian at a given order.
Contributions of non-renormalizable interactions in physical quantities are expected to be suppressed by
scales much larger than the considered energies. However, usually it is meant that this assumption does not apply to
EFTs  with non-Abelian massive vector bosons unless the masses are generated by spontaneous breaking of a gauge symmetry.
This is because the limit of the vanishing
mass of non-Abelian vector bosons generates divergent results (containing inverse powers of the small mass)
at each order of perturbation theory \cite{Boulware:1970zc,Slavnov:1970tk}.
In EFT with massive vector bosons
without spontaneous symmetry breaking the inverse powers of the vector boson mass enter the renormalization of
couplings with negative mass dimensions and therefore it appears that these couplings cannot be suppressed by powers of some large scale.
A closely related well-known problem with non-Abelian massive vector bosons is that the perturbative expressions of cross
sections involving longitudinal components of vector bosons grow rapidly at high energies and violate the unitary bound.
This might be the result of internal inconsistency of such a theory or merely indicate the failure of perturbation theory due to
the presence of resonances (or bound states).
Interesting new  results of non-perturbative studies of the massive
Yang-Mills theory on the lattice have been recently reported in Refs.~\cite{Ferrari:2013lja,DellaMorte:2013yca}.
In particular, indications of the presence of a scalar state in the spectrum of particles have been observed.
In  appendix~B of this paper we provide with a simple ''toy model'' example of qualitatively different behaviors of
perturbative and non-perturbative expressions in the limit of vanishing mass parameter $M\to 0$, which we believe is a good demonstration of the possible behavior in a quantum field theory of massive vector bosons.

In Ref.~\cite{Gegelia:2011fp} it has been shown that an EFT of massive Yang-Mills vector bosons
interacting with fermions is renormalizable in the sense of EFT and it was argued, using the results of Ref.~\cite{Vainshtein:1971ip},
that the vanishing vector boson mass limit exists non-perturbatively.
If this is the case then the divergent high-energy behavior of cross sections
must be indeed an artifact of the presence of resonances (or bound states) and consequently one
would expect that the perturbative expressions of physical quantities characterizing the static properties of particles, which do not depend on any kinematical variables,
should not diverge in the limit of vanishing vector boson mass.

	In this work we calculate two-loop corrections to the pole mass of the vector boson and the pole masses and the magnetic moments of fermions  in the framework of an EFT of massive Yang-Mills vector bosons interacting with fermions of Ref.~\cite{Gegelia:2011fp}. The pole mass of the vector boson vanishes, and the other considered quantities prove to be finite in the limit of vanishing vector boson mass parameter of the Lagrangian. We also discuss some implications of the obtained results.

\section{Effective Lagrangian }

    We consider an EFT of the massive Yang-Mills vector fields
interacting with fermions given by the following Lagrangian \cite{Gegelia:2011fp}
(for a general discussion of vector meson Lagrangians, see e.g. Ref.~\cite{Meissner:1987ge})
\begin{eqnarray}
{\cal L} & = & -\frac{1}{4}\,\sum_a\, G^{a \mu\nu}
G^a_{\mu\nu}+\frac{M^2}{2}\,\sum_a B^\mu_a B_{a \mu} +
\frac{g^2 \theta}{64 \pi^2}\,\epsilon^{\mu\nu\alpha\beta} G^a_{\mu\nu}
G^a_{\alpha\beta}\nonumber\\
&& + \sum_{q,j} \bar \psi_q^j \left( i
\partial \hspace{-.6em}/\hspace{.1em}-m_q\right)
\psi_q^j
 +g\sum_{q,i,k,a} \bar \psi_q^i\gamma_\mu t^a_{ik}\psi_q^k B_a^\mu+{\cal
L}_1\,,
 \label{lagrangianQCD01}
\end{eqnarray}
where $G^a_{\mu\nu}=\partial_\mu B^a_\nu-\partial_\nu B^a_\mu + g
f^{abc} B_\mu^bB_\nu^c$, $t^a$ and $f^{abc}$ are the generators and the totally
antisymmetric structure constants of the $SU(N)$ group, respectively.
The index $q=1,\ldots, N_f$ enumerates the different types (flavours) of fermions.
	The ${\cal L}_1$ part of the effective Lagrangian contains all possible Lorentz- and
gauge-invariant terms (an infinite number of them) with
coupling constants with negative mass dimensions. This guarantees that the considered EFT is
renormalizable to all orders in loop expansion \cite{Gegelia:2011fp}.
It is assumed that contributions to physical quantities of renormalized
coupling constants with negative mass dimensions
are suppressed by powers of some large scale. Note that this assumption is only justified if the $M\to 0$ limit leads indeed
to finite physical quantities in the non-perturbative regime.

	The generating functional of Green's functions is given by
\begin{equation}
Z[J^a_\mu,\xi,\bar\xi] = \int {\cal D} B\,{\cal D} \psi\,{\cal D}
\bar\psi\,\exp \left\{ i \int d^4 x \,\left[ {\cal
L}(x)+J^{a\mu}B^a_\mu +\xi \bar\psi + \bar \xi\psi
\right]\right\}, \label{GFFinal0}
\end{equation}
where the indices are suppressed for fermion fields.
It is straightforward to obtain the standard Feynmann rules using Eq.~(\ref{GFFinal0}).
The propagator of the vector boson has the form
\begin{equation}
i D^{\mu\nu}(p)=-i\, \frac{g^{\mu\nu}-p^\mu p^\nu/M^2}{p^2-M^2+i\,\epsilon}.
\label{SF}
\end{equation}
The $g^{\mu\nu}$ part of the propagator can be interpreted as an infrared regularized version of the Feynman-gauge
propagator  of the vector boson in the standard massless Yang-Mills theory. Therefore the $M\to 0$ limit of contributions in physical
quantities generated by this term of the propagator leads to finite
results if the corresponding expressions are infrared finite in the massless Yang-Mills theory.
This property will be used in later calculations.

Below using dimensional regularization and the modified minimal subtraction scheme ($\overline{MS}$) (see e.g. Ref.~\cite{Collins:1984xc}) we calculate
two-loop corrections to the pole mass of the vector boson and the pole masses
and the magnetic moments of fermions and consider the $M\to 0$ limit of these quantities.

\section{Pole mass of the vector boson}

The self-energy of the vector boson is defined as a sum of all one-particle irreducible diagrams contributing in the two point
function. For the considered effective field theory it has a transverse structure \cite{Gegelia:2011fp}:
\begin{equation}
i \delta^{ab} \Pi_{\mu\nu} =i \delta^{ab} \left(g_{\mu\nu} p^2-p_\mu p_\nu\right) \Pi(p^2).
\label{BSE}
\end{equation}
The corresponding dressed propagator of the vector boson has the form
\begin{equation}
i S_{\mu\nu}^{ab}(p) = - i\,\delta^{ab} \frac{g_{\mu\nu} -p_\mu p_\nu \frac{1-\Pi(p^2)}{M^2}} {p^2-M^2-p^2 \Pi(p^2)+i\epsilon}.
\label{DBPR}
\end{equation}
The pole mass of the vector boson $M_R$ is obtained by solving the equation
\begin{equation}
M_R^2-M^2-M_R^2\Pi(M_R^2)=0.
\label{VBPME}
\end{equation}

\begin{figure}
\epsfig{file=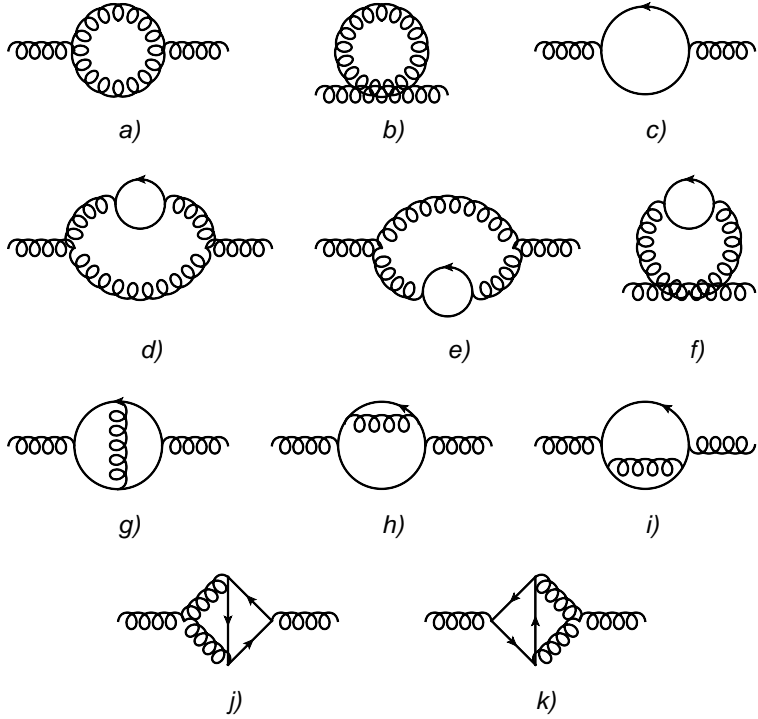, width=10cm}
\caption[]{\label{VSE} One- and two-loop contributions to the self-energy of the vector boson. Solid and wiggly lines correspond
to the fermion and the vector boson, respectively.}
\end{figure}

\noindent
The solution to Eq.~(\ref{VBPME}) can be obtained perturbatively order-by-order by parameterizing the square of the pole mass as
\begin{equation}
M_R^2=M^2+\sum_i \delta M_i^2,
\label{MRpar}
\end{equation}
where the summation over $i$ refers to different orders in coupling constants. Here we calculate the contributions of one- and two-loop
diagrams shown in Fig.~\ref{VSE}, which are generated by interaction terms explicitly displayed in Eq.~(\ref{lagrangianQCD01}), i.e. interaction
terms depending on the coupling constant $g$ only.

The $g^2$ and $g^4$ contributions of the one-loop self-energy diagrams and the $g^4$ order contributions of two-loop diagrams
in the pole mass of the vector boson are specified in appendix~A (in case of two-loop diagrams we give only expressions of those terms, which involve inverse powers of $M^2$). All these contributions vanish in the $M\to 0$ limit, i.e. the inverse powers of $M^2$ in $p^\mu p^\nu$ parts of the vector boson propagator do not lead to any singularities in this case.

\section{Pole masses and magnetic moments of fermions}

\begin{figure}
\epsfig{file=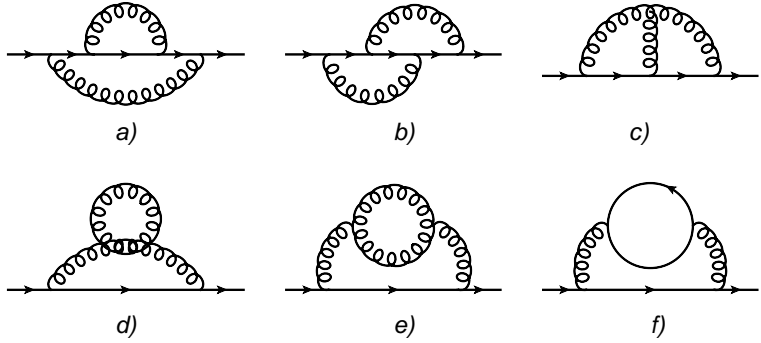, width=10cm}
\caption[]{\label{mass} Two-loop contributions to the fermion self-energy. Solid and wiggly lines correspond
to the fermion and the vector boson, respectively.}
\end{figure}

The pole mass of the $i$-th  fermion $m_{i p}$ is found by solving the equation
\begin{equation}
\biggl[p\hspace{-.4em}/\hspace{.1em}-m_i-\Sigma(p\hspace{-.4em}/\hspace{.1em})\biggr]|_{p\hspace{-.4em}/\hspace{.1em}=m_{ip}}=0,
\label{dressedFpropagator}
\end{equation}
where $m_i$ is the renormalized mass of the $i$-th fermion and $\Sigma(p\hspace{-.4em}/\hspace{.1em})$ is the fermion self-energy.

The two-loop diagrams contributing to the fermion self-energy are shown in Fig.~\ref{mass}.
The fermion pole mass at two-loop order is gauge-independent and infrared finite
in massless Yang-Mills theory \cite{Gray:1990yh,Fleischer:1998dw}, therefore the contribution of products of $g^{\mu\nu}$
parts of vector boson propagators of diagrams of Fig.~\ref{mass} are finite in the $M\to 0$ limit. The contribution of the diagram f) is also finite in that limit. Diagram a) and the part of the diagram b), proportional to $C_F^2$, together with the contribution of the one-loop diagram give a finite result in $M\to 0$ limit, as expected because the same diagrams contribute in the Abelian case, where the massless limit coincides with the results of the massless theory \cite{Boulware:1970zc}.
The remaining contributions of all two-loop diagrams, i.e. those pieces which are generated by parts of vector boson propagators, containing inverse powers of $M$, sum up to the following expression:
\begin{eqnarray}
m_{i p,2} &=& -\frac{g^4 C_A C_F}{2\,(2 \pi)^{2 n}}\int \frac{d^nk_1 d^nk_2}{[k_1^2-M^2]^2
   [k_2^2-M^2][(k_1+p)^2-m_i^2]}\nonumber\\
&& \times
   \left\{ k_2\hspace{-0.9em}/\hspace{.1em} \ (m_i+k_1\hspace{-0.9em}/\hspace{.1em}
   +p\hspace{-0.4em}/\hspace{.1em}) k_2\hspace{-0.9em}/\hspace{.1em} \, \left[
   \frac{M^2}{(k_1+k_2)^2-M^2}-\frac{2}{M^2}\right]+2\,\gamma
   ^{\beta } (m_i+k_1\hspace{-0.9em}/\hspace{.1em}
   +p\hspace{-0.4em}/\hspace{.1em}) \gamma
   ^{\beta }\right\},
\label{mass2loop}
\end{eqnarray}
where $C_A=N$ and $C_F=(N^2-1)/(2 N)$. Using the method of dimensional counting of Ref.~\cite{Gegelia:1994zz} it is easily verified that
$m_{i p,2}$ is finite in $M\to 0$ limit. Note that the contributions of separate diagrams are divergent in the considered limit,
i.e. the finite result is due to non-trivial cancelations among various diagrams. In particular, while the terms proportional to $1/M^8$ and $1/M^6$ factors generated by vector boson propagators vanish individually for diagrams, the terms proportional to $1/M^4$, $1/M^2$ and $\ln M$ cancel only by adding contributions of all two-loop diagrams.

\begin{figure}
\epsfig{file=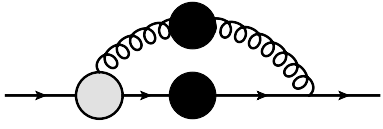, width=5cm}
\caption[]{\label{massDS} Dyson-Schwinger representation of the fermion self-energy. Solid and wiggly lines correspond
to fermions and vector bosons, respectively. The circles stand for the corresponding dressed quantities.}
\end{figure}

By using the Dyson-Schwinger representation of the self-energy of the fermion (see e.g \cite{Roberts:1994dr})
shown in Fig.~\ref{massDS} we can analyze all diagrams contributing in fermion self-energy which are generated by interaction terms explicitly shown in the Lagrangian Eq.~(\ref{lagrangianQCD01}), i.e. diagrams depending only on one coupling constant $g$. From the
Slavlov-Taylor identities of Ref.~\cite{Gegelia:2011fp} it follows straightforwardly that the $p^\mu p^\nu$ part of the dressed vector boson propagator does not contribute in the fermion pole mass. This observation also supports the expectation that the physical quantities are finite in $M\to 0$ limit.

\medskip

\begin{figure}
\epsfig{file=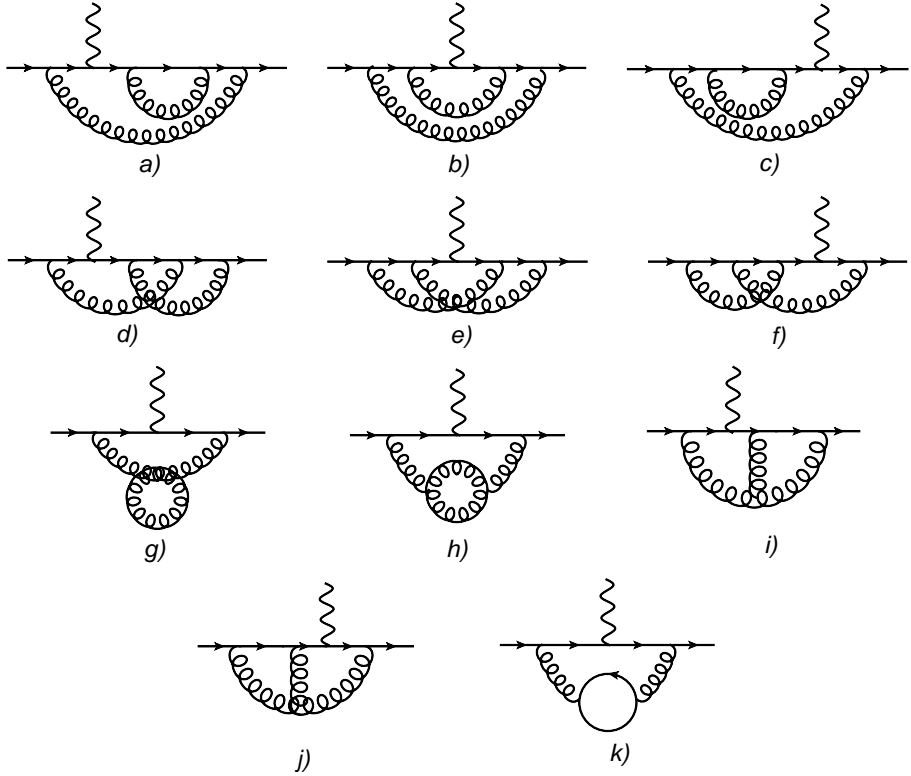, width=12cm}
\caption[]{\label{fmm} Two-loop contributions to the magnetic moment of the fermion. Solid, wavy and wiggly lines correspond
to the fermion, electromagnetic current and the vector boson, respectively.}
\end{figure}

Next we consider the magnetic moment of $i$-th fermion defined in standard way through the matrix element of the
electromagnetic current. The two-loop diagrams contributing in the magnetic moment of the fermion are shown in Fig.~\ref{fmm}.
It is easily checked that the result of the diagram k) is finite in $M\to 0$ limit. The result of diagram a) together with the
part of diagram b) proportional to $C_F^2$, and the contribution of the one-loop diagram multiplied with the wave
function renormalization constant also lead to a finite result in $M\to 0$ limit. This was to be expected, as the same diagrams contribute in the Abelian case, where it is known that the massless limit coincides with the results of the massless theory \cite{Boulware:1970zc}.
Next, the two-loop contribution to the magnetic moment of a fermion (quark) is gauge-independent and
infrared finite in the standard Yang-Mills theory (QCD) \cite{Japaridze:1998tt}. Therefore the contributions
of products of $g^{\mu\nu}$ parts of the vector boson propagators generate a result which is finite in the $M\to 0$ limit. The remaining contributions of all two-loop diagrams, i.e. those which are generated by parts of vector boson propagators  containing inverse powers of $M$, sum up to the following
expression:
\begin{eqnarray}
\kappa &=& \frac{e\,g^4 C_A C_F}{2\,(2 \pi)^{2 n}} \int \frac{d^nk_1 d^nk_2}{[k_1^2-M^2]^2 [k_2^2-M^2][(k_1+p_i )^2-m_i^2][(k_1+p_f)^2-m_i^2]}
\nonumber\\
&& \times \Biggl\{\left(
   \frac{1}{(k_1+k_2)^2-M^2}-\frac{2}{M^2}\right)
   k_2\hspace{-0.9em}/\hspace{.1em} \ (m_i+k_1\hspace{-0.9em}/\hspace{.1em}
   +p_f\hspace{-0.88em}/\hspace{.1em}\,) \gamma ^{\mu }
   (m_i+k_1\hspace{-0.9em}/\hspace{.1em}
   +p_i\hspace{-0.7em}/\hspace{.1em}) k_2\hspace{-0.9em}/\hspace{.1em} \nonumber\\
&& +2 \gamma ^{\beta
   } (m_i+k_1\hspace{-0.9em}/\hspace{.1em}
   +p_f\hspace{-0.88em}/\hspace{.1em}\,) \gamma ^{\mu }
   (m_i+k_1\hspace{-0.9em}/\hspace{.1em}
   +p_i\hspace{-0.7em}/\hspace{.1em})\gamma
   ^{\beta }\Biggr\}.
\end{eqnarray}
Using the method of dimensional counting of Ref.~\cite{Gegelia:1994zz} it is easily seen that the only part of $\kappa$
which might diverge in the limit $M\to 0$ is
\begin{eqnarray}
\kappa_d &=& \frac{e\,g^4 C_A C_F}{2\,(2 \pi)^{2 n}} \int \frac{d^nk_1 d^nk_2}{[k_1^2-M^2]^2 [k_2^2-M^2][(k_1+p_i )^2-m_i^2][(k_1+p_f)^2-m_i^2]}
\nonumber\\
&& \times \Biggl\{\left(
   \frac{1}{(k_1+k_2)^2-M^2}-\frac{2}{M^2}\right)
   k_2\hspace{-0.9em}/\hspace{.1em} \ (m_i+p_f\hspace{-0.88em}/\hspace{.1em}\,) \gamma ^{\mu }
   (m_i+p_i\hspace{-0.7em}/\hspace{.1em}) k_2\hspace{-0.9em}/\hspace{.1em} \nonumber\\
&& +2 \gamma ^{\beta
   } (m_i+p_f\hspace{-0.88em}/\hspace{.1em}\,) \gamma ^{\mu }
   (m_i+p_i\hspace{-0.7em}/\hspace{.1em})\gamma
   ^{\beta }\Biggr\}.
\end{eqnarray}
Sandwiching $\kappa_d$ between Dirac spinors $\bar u(p_f)$ and $u(p_i)$ we obtain:
\begin{eqnarray}
\bar u (p_f) \kappa_d u (p_i) & = & \frac{2 e\,g^4 C_A C_F }{(2 \pi)^{2 n}} \int \frac{d^nk_1 d^nk_2 \,\bar u (p_f) \gamma
   ^{\mu } u (p_i)}{[k_1^2-M^2]^2 [k_2^2-M^2][(k_1+p_i )^2-m_i^2][(k_1+p_f)^2-m_i^2]}
\nonumber\\
&& \times \Biggl\{\left(
   \frac{1}{(k_1+k_2)^2-M^2}-\frac{2}{M^2}\right)
   k_2\cdot p_i \, k_2\cdot p_f  +2 p_f\cdot p_i \Biggr\}.
\label{kappas}
\end{eqnarray}
It is easily seen from Eq.~(\ref{kappas}) that $\kappa_d$ does not contribute in the magnetic moment,
and hence the whole two-loop correction to the magnetic
moment of the fermion is finite in $M\to 0$ limit. Again the finite result is due to the non-trivial cancelation of severely
divergent contributions of various diagrams.

\medskip

\section{summary and discussion}

In this work we calculated two-loop corrections to the vector boson pole mass and the pole masses and the magnetic moments of fermions  in the framework of an EFT of massive Yang-Mills vector bosons interacting with fermions of Ref.~\cite{Gegelia:2011fp}. All these quantities are finite (non-divergent) when the vector boson mass parameter of the Lagrangian is taken to zero.
In particular, the one and two-loop order corrections to the pole mass of the vector boson vanish in $M\to 0$ limit.
We expect that the non-trivial cancelations of divergent contributions of various diagrams in these physical quantities is not accidental
so that the physical quantities characterizing
the static properties of particles of the spectrum are finite to all orders in the limit of vanishing vector boson mass.
It is rather difficult to tell if the vanishing mass limit reproduces the corresponding quantities of the massless theory or  the
van Dam-Veltman-Zakharov (vDVZ) discontinuity \cite{vanDam:1970vg,Zakharov:1970cc} persists. This is because our results
of static characteristics are expressed in terms of parameters in $\overline{MS}$ scheme, which are not physical quantities and it is not clear how they are related to corresponding parameters of the massless theory. To find out whether  the discontinuity appears one needs to express physical quantities in terms of other physical quantities and after consider the massless limit. More physical quantities need to be calculated to answer this question. Investigation of this issue is the subject of a future
study.
The severe divergences in scattering amplitudes we interpret as an evidence of a
non-trivial analytic structure of these amplitudes due to the presence of resonances (or bound states).
Demanding self-consistency of an EFT of massive vector bosons interacting with fermions one is necessarily lead at leading order
to the Lagrangian of massive Yang-Mills theory \cite{Djukanovic:2010tb,Gegelia:2012yq}. However the applicability of this theory is restricted by the
position of the above mentioned resonances (bound states).
A consistent inclusion of dynamical fields corresponding to these resonances in the effective Lagrangian extends the region of applicability of the theory by removing the divergences appearing when $M\to 0$ limit is taken. As these divergences are caused by the longitudinal part of the vector boson propagator, the resonance(s) which have to be included as dynamical degree(s) of freedom could be realized as a scalar field(s).  By demanding
tree-order unitarity of the $S$ matrix for vector bosons interacting with fermion and scalar fields one is necessarily lead to
the gauge-invariant theory with spontaneous symmetry breaking \cite{LlewellynSmith:1973ey,Cornwall:1973tb,Cornwall:1974km,Joglekar:1973hh}.


\section*{Acknowledgments}

J.~G. thanks J.~Collins for discussions.
This work was supported in part by Georgian Shota Rustaveli National
Science Foundation (grant 11/31), by the EPOS  network of the European Community
Research Infrastructure Integrating Activity ``Study of Strongly Interacting
Matter'' (HadronPhysics3, Grant No. 283286), and by the DFG (SFB/TR 16,
``Subnuclear Structure of Matter'').

\appendix

\section{Vector-boson self-energy}

The contributions of one-loop diagrams a), b) and c) to the vector boson self-energy function $\Pi(p^2)$ have the form:
\begin{eqnarray}
\Pi_{a+b} &=& \frac{g^2 C_A}{1152 \pi ^2 M^4 p^2}  \biggl[3 \left(48 M^6+68 M^4 p^2-16 M^2 p^4-p^6\right)
   B_0\left(p^2,M^2,M^2\right) \nonumber\\
&+& 6 \left(-24 M^4+8 M^2 p^2+p^4\right)
   A_0\left(M^2\right)+144 M^6-48 M^4 p^2+20 M^2 p^4-2
   p^6\biggr],\nonumber\\
\Pi_{c} &=& \sum_{i=1}^{N_f} \frac{g^2  \biggl\{p^2 \left[1-3
   B_0\left(p^2,m_i^2,m_i^2\right)\right]-6 m_i^2
   \left[B_0\left(p^2,m_i^2,m_i^2\right)+1\right]+6
   A_0\left(m_i^2\right)\biggr\}}{72 \pi ^2 p^2}.
\label{BSE12}
\end{eqnarray}
Here, the loop functions $A_0$ and $B_0$ are defined as follows:
\begin{eqnarray}
A_0(m^2) & = & \frac{(2 \pi \mu)^{4-n}}{i\,\pi^2}\,\int \frac{
d^nk}{k^2-m^2}\,, \nonumber\\
B_0(p^2,m_1^2,m_2^2) & = & \frac{(2 \pi \mu)^{4-n}}{i\,\pi^2}\,\int
\frac{d^nk}{\left[k^2-m_1^2\right]\left[(p+k)^2-m_2^2\right]}\,,
\label{oneandtwoPF}
\end{eqnarray}
where $n$ is the space-time dimension.
The $g^2$ and $g^4$ contributions of the above expressions in the pole mass of the vector boson read:
\begin{eqnarray}
\delta M_{1loop}^2 &=& \frac{g^2 C_A \biggl\{M^2 \left[99
   B_0\left(M^2,M^2,M^2\right)+38\right]-30
   A_0\left(M^2\right)\biggr\}}{384 \pi ^2} \nonumber\\
&+& \sum_{i=1}^{N_f}\frac{g^2}{72 \pi ^2} \left[-3 \left(2 m_i^2+M^2\right)
   B_0\left(M^2,m_i^2,m_i^2\right)+6 A_0\left(m_i^2\right)-6
   m_i^2+M^2\right]\nonumber\\
&+& \frac{g^4 C_A^2}{442368 \pi ^4 M^2} \biggl\{M^2 \left[99
   B_0\left(M^2,M^2,M^2\right)+38\right]-30
   A_0\left(M^2\right)\biggr\}\nonumber\\
&\times& \biggl\{M^2 \left[99
   B_0\left(M^2,M^2,M^2\right)+66 \sqrt{3} \pi -311\right]+60
   A_0\left(M^2\right)\biggr\}\nonumber\\
&+& \sum_{i=1}^{N_f} \frac{g^4 C_A}{82944 \pi ^4 M^2 \left(M^2-4 m_i^2\right)}
\Biggl\{M^2 \biggl[6 A_0\left(m_i^2\right) \biggl(198
   B_0\left(M^2,M^2,M^2\right) \left(m_i^2+2 M^2\right) \nonumber\\
&+& 66 \sqrt{3}
   \pi  \left(M^2-4 m_i^2\right)+1472 m_i^2-197 M^2\biggr) \nonumber\\
&+& 2 M^2
   m_i^2 \biggl(3 \left(396 B_0\left(M^2,M^2,M^2\right)+66 \sqrt{3}
   \pi -197\right) B_0\left(M^2,m_i^2,m_i^2\right) \nonumber\\
&-& 1089
   B_0\left(M^2,M^2,M^2\right)-330 \sqrt{3} \pi +1327\biggr) \nonumber\\
&+& 12
   m_i^4 \left(-99 B_0\left(M^2,M^2,M^2\right)+132 \sqrt{3} \pi
   -736\right) \left(B_0\left(M^2,m_i^2,m_i^2\right)+1\right) \nonumber\\
&+& M^4
   \biggl(-3 \left(396 B_0\left(M^2,M^2,M^2\right)+66 \sqrt{3} \pi
   -197\right) B_0\left(M^2,m_i^2,m_i^2\right) \nonumber\\
&+& 1287
   B_0\left(M^2,M^2,M^2\right)+66 \sqrt{3} \pi +145\biggr)\biggr] \nonumber\\
&+& 30
   A_0\left(M^2\right) \biggl(3 \left(-2 M^2 m_i^2+28
   m_i^4+M^4\right) B_0\left(M^2,m_i^2,m_i^2\right) \nonumber\\
&-& 6 \left(14
   m_i^2+M^2\right) A_0\left(m_i^2\right)-8 M^2 m_i^2+84 m_i^4-10
   M^4\biggr)\Biggr\}\nonumber\\
&+& \sum_{i,j=1}^{N_f}\frac{g^4}{5184 \pi ^4 M^2
   \left(M^2-4 m_i^2\right)}   \biggl[3 \left(-2 M^2 m_i^2+4 m_i^4+M^4\right)
   B_0\left(M^2,m_i^2,m_i^2\right) \nonumber\\
&-& 6 \left(2 m_i^2+M^2\right)
   A_0\left(m_i^2\right)+4 M^2 m_i^2+12 m_i^4-4 M^4\biggr] \nonumber\\
&\times& \biggl[3
   \left(2 m_j^2+M^2\right) B_0\left(M^2,m_j^2,m_j^2\right)-6
   A_0\left(m_i^2\right)+6 m_j^2-M^2\biggr].
\label{polemass1loop}
\end{eqnarray}

Below we give the non-vanishing contributions of those parts of the two-loop diagrams in the vector boson self-energy
which contain inverse powers of $M^2$ (generated by the $p^\mu p^\nu$ parts of vector boson propagators)
\begin{eqnarray}
\Pi^{\mu\nu}_d &=& \Pi^{\mu\nu}_e=-\frac{g^4 C_A}{4 M^2}  \sum_{i=1}^{N_f}\int\frac{d^nk_1d^nk_2}{(2 \pi) ^{2n} }
   \frac{\gamma ^{\nu } (m_i+
   k_1\hspace{-0.88em}/\hspace{.1em})\gamma ^{\mu }(m_i+
   k_1\hspace{-0.88em}/\hspace{.1em}+k_2\hspace{-0.88em}/\hspace{.1em}+p\hspace{-0.45em}/\hspace{.1em})}
   {[k_2^2-M^2][k_1^2-m_i^2][(k_1+k_2+p)^2-m_i^2]},\nonumber\\
\Pi^{\mu\nu}_g &=& \frac{g^4 (C_A-2 C_F)}{2
   M^2} \int \frac{d^nk_1 d^nk_2}{(2 \pi) ^{2n} }\, {\rm Tr} \Biggl\{\frac{\gamma ^{\nu
   } (m_i+k_1 \hspace{-0.88em}/\hspace{.1em}+ p\hspace{-0.45em}/\hspace{.1em}) \gamma ^{\mu
   }(m_i+k_1\hspace{-0.88em}/\hspace{.1em})}
   {\left[k_2^2-M^2\right]\left[k_1^2-m_i^2\right]\left[(k_1+p)^2-m_i^2\right]}\nonumber\\
&-& \frac{\gamma ^{\nu }(m_i+k_1\hspace{-0.88em}/\hspace{.1em}+
k_2\hspace{-0.88em}/\hspace{.1em}+p\hspace{-0.45em}/\hspace{.1em}) \gamma
   ^{\mu } (m_i+k_1\hspace{-0.88em}/\hspace{.1em})}
   {\left[k_2^2-M^2\right]\left[k_1^2-m_i^2\right]\left[(k_1+k_2+p)^2-m_i^2\right]}\Biggr\},\nonumber\\
\Pi^{\mu\nu}_h &=& \Pi^{\mu\nu}_i=-\frac{g^4 C_F}{2 M^2} \sum_{i=1}^{N_f}\int\frac{d^nk_1d^nk_2}{(2 \pi) ^{2n} }
   \Biggl\{\frac{\gamma ^{\nu
   }(m_i+k_1\hspace{-0.88em}/\hspace{.1em}+k_2\hspace{-0.88em}/\hspace{.1em}
   +p\hspace{-0.4em}/\hspace{.1em})\gamma ^{\mu }(m_i+k_1\hspace{-0.88em}/\hspace{.1em})}{[k_2^2-M^2][k_1^2-m_i^2][(k_1+k_2+p)^2-m_i^2]}
\nonumber\\
&-& \frac{\gamma ^{\nu
   }(m_i+k_1\hspace{-0.88em}/\hspace{.1em}+p\hspace{-0.45em}/\hspace{.1em})\gamma ^{\mu
   }(m_i+k_1\hspace{-0.88em}/\hspace{.1em})}{[k_2^2-M^2][k_1^2-m_i^2][(k_1+p)^2-m_i^2]}\Biggr\},\nonumber\\
\Pi^{\mu\nu}_j &=& \Pi^{\mu\nu}_k =\frac{g^4 C_A }{
   4 M^2} \sum_{i=1}^{N_f}\int\frac{d^nk_1d^nk_2}{(2 \pi) ^{2n} } \Biggl\{ \frac{\gamma ^{\nu } (m_i+k_1\hspace{-0.88em}/\hspace{.1em}+k_2\hspace{-0.88em}/\hspace{.1em}
   +p\hspace{-0.45em}/\hspace{.1em}) \gamma
   ^{\mu } (m_i+k_1\hspace{-0.88em}/\hspace{.1em})}{[k_2^2-M^2][k_1^2-m_i^2][(k_1+k_2+p)^2-m_i^2]}\nonumber\\
&-& \frac{2 \gamma ^{\nu }
   (m_i+ k_1\hspace{-0.88em}/\hspace{.1em}+p\hspace{-0.45em}/\hspace{.1em}) \gamma ^{\mu }(m_i+
   k_1\hspace{-0.88em}/\hspace{.1em}) }{[k_1^2-m_i^2][k_2^2-M^2][(k_1+p)^2-m_i^2]}\nonumber\\
&+& \frac{2
   k_2^{\mu } \gamma ^{\nu
   } (m_i+k_1\hspace{-0.88em}/\hspace{.1em}+p\hspace{-0.45em}/\hspace{.1em})
   k_2\hspace{-0.88em}/\hspace{.1em} \ (m_i+k_1\hspace{-0.88em}/\hspace{.1em})}
   {[k_2^2-M^2][k_1^2-m_i^2][(k_2+p)^2-M^2][(k_1+p)^2-m_i^2]}\nonumber\\
&+&
   \frac{\gamma ^{\nu }(m_i+
   k_1 \hspace{-0.88em}/\hspace{.1em} +p\hspace{-0.45em}/\hspace{.1em})\gamma ^{\mu }(m_i+
   k_1\hspace{-0.88em}/\hspace{.1em}+k_2\hspace{-0.88em}/\hspace{.1em}+p\hspace{-0.45em}/\hspace{.1em})}
   {[(k_2+p)^2-M^2][(k_1+p)^2-m_i^2][(k_1+k_2+p)^2-m_i^2]}\Biggr\}\nonumber\\
&-& \frac{g^4 C_A p^2}{4 M^4}  \sum_{i=1}^{N_f}\int\frac{d^nk_1d^nk_2}{(2 \pi) ^{2n} }
   \frac{k_2^{\mu } \gamma ^{\nu } (m_i+
   k_1\hspace{-0.88em}/\hspace{.1em}+p\hspace{-0.45em}/\hspace{.1em}) k_2\hspace{-0.88em}/\hspace{.1em}\
   (m_i+k_1\hspace{-0.88em}/\hspace{.1em})  }{[k_2^2-M^2][k_1^2-m_i^2][(k_2+p)^2-M^2][(k_1+p)^2-m_i^2]},\nonumber\\
\Pi^{\mu\nu}_e &=& 0 .
\label{2loopD}
\end{eqnarray}
Adding the above results of two-loop diagrams after some simplification we finally obtain for the contribution in the vector boson mass:
\begin{eqnarray}
\delta M_{2loop}^2 &=&\frac{g^4 C_A}{2048 \pi ^4 M^2 (n-1)^2} \left[2 (4 n-7) A_0\left(M^2\right)-9 M^2
   B_0\left(M^2,M^2,M^2\right)\right]\nonumber\\
&\times&  \left[\left(4 m_i^2+M^2
   (n-2)\right) B_0\left(M^2,m_i^2,m_i^2\right)-2 (n-2)
   A_0\left(m_i^2\right)\right].
\label{2loopmass}
\end{eqnarray}
Using the expressions of loop functions
\begin{eqnarray}
A_0(m^2)&=& -\frac{2 m^2}{n-4}-2 m^2 \ln\frac{m}{\mu}-\gamma_E
   m^2+m^2+m^2 \ln(4 \pi ),\nonumber\\
B_0(p^2,m_1^2,m_2^2) & = & -\frac{2}{n-4}-\frac{2
   m_1 m_2}{p^2}\, \sqrt{1-\frac{\left(m_1^2+m_2^2-p^2\right){}^2}{4 m_1^2 m_2^2}} \cos
   ^{-1}\left(\frac{m_1^2+m_2^2-p^2}{2 m_1 m_2}\right)\nonumber\\
&-& 2 \ln
   \frac{m_2}{\mu}-\gamma +2+\ln (4 \pi )-\frac{1}{p^2}\,\left(m_1^2-m_2^2+p^2\right) \ln \frac{m_1}{m_2},
\label{AandB}
\end{eqnarray}
where $\gamma_E$ is the Euler-gamma, we obtain that $\delta M_{1loop}^2$ and $\delta M_{2loop}^2$ both vanish in the $M\to 0$ limit.

\section{Toy model}

Below we give a simple example that demonstrates the qualitatively different behavior of
perturbatively and non-pertubatively calculated '``physical'' quantities in $M\to 0$ limit.
We consider numerical integrals, ``Green functions'' of the  ``quantum field theory in zero dimensions''
with $x$ and $y$ ``fields'' with ``masses'' $M$ and $m$
\begin{equation}
G_N=\frac{\int_0^\infty dx \int_0^\infty y^2 dy \, y^{N} e^{-M^2 x^2-m^2 y^2-\lambda x y^2}}{\int_0^\infty dx \int_0^\infty y^2 dy \,e^{-M^2 x^2-m^2 y^2-\lambda x y^2}}.
\label{grf}
\end{equation}
For the case of massless $x$ field the ''Green functions'' are given as
\begin{equation}
G_N^0=\frac{\int_0^\infty dx \int_0^\infty y^2 dy \, y^{N} e^{-m^2 y^2-\lambda x y^2}}{\int_0^\infty dx \int_0^\infty y^2 dy \,e^{-m^2 y^2-\lambda x y^2}}.
\label{grf0}
\end{equation}

A perturbative calculation of $G_2$ as a power series in $\lambda$ gives
\begin{equation}
G_2 = \frac{3}{2 m^2}-\frac{3 \lambda }{2 \sqrt{\pi } m^4 M}+\frac{3 (5 \pi -6) \lambda ^2}{8 \pi  m^6 M^2}
-\frac{3 (36+25 \pi ) \lambda ^3}{32 \pi ^{3/2}
   m^8 M^3} +\cdots .
\label{GNpert}
\end{equation}
This demonstrates that the $M\to 0$ limit does not exist in perturbation theory.
On the other hand, integrating Eqs.~(\ref{grf}) and (\ref{grf0})
over $x$ and $y$ we can easily verify that
\begin{equation}
\lim_{M\to 0} G_N \to G_N^0.
\label{limMto0}
\end{equation}
Thus, while the perturbation theory suggests that the $M\to 0$ limit does not exist, this limit is well defined
non-perturbatively and coincides with the massless ``theory''.

\medskip

The effective ``Lagrangian'' of the field $y$ is obtained by integrating over $x$:
\begin{eqnarray}
L_{\rm eff} & = & 
-m^2 y^2+\frac{y^4
   \lambda ^2}{4 M^2}+\ln \left[\text{erfc}\left(\frac{y^2 \lambda
   }{2 M}\right)\right]\nonumber\\
   & = & \ln \frac{\sqrt{\pi}}{2 M} -\left(m^2 +\frac{\lambda }{\sqrt{\pi}\,M}\right)\,y^2
   +\frac{(\pi -2) y^4 \lambda ^2}{4 \pi  M^2}+\frac{(\pi -4) y^6 \lambda ^3}{12 \pi ^{3/2}
   M^3}+\cdots .
\label{Leff}
\end{eqnarray}
Further, the coupling constants of the effective Lagrangian of Eq.~(\ref{Leff}) contain inverse powers of $M$.
These couplings would diverge in $M\to 0$ limit, however these divergences are misleading.
The actual effective lagrangian for $M\to 0$ is obtained by integrating over $x$ in Eq.~(\ref{grf0}) and has the form
\begin{eqnarray}
L_{\rm eff} & = & -m^2 y^2+\ln (\lambda y^2).
\label{Leff0}
\end{eqnarray}
Note that for massless $x$ ``field'' perturbation theory generates divergent terms, which are not relevant for our purposes. These could be e.g. interpreted as analogues of infrared divergences.



\end{document}